\documentclass[pdflatex,sn-vancouver-ay]{sn-jnl} % Vancouver Author Year Reference Style
\usepackage{graphicx}%
\usepackage{multirow}%
\usepackage{amsmath,amssymb,amsfonts}%
\usepackage{amsthm}%
\usepackage{mathrsfs}%
\usepackage[title]{appendix}%
\usepackage{xcolor}%
\usepackage{textcomp}%
\usepackage{manyfoot}%
\usepackage{booktabs}%
\usepackage{algorithm}%
\usepackage{algorithmicx}%
\usepackage{algpseudocode}%
\usepackage{listings}%

\usepackage{hyperref}
\theoremstyle{thmstyleone}%
\theoremstyle{thmstyletwo}%

\theoremstyle{thmstylethree}%

\begin{document}

\title{Bridging the Dual Nature: How Integrated Explanations Enhance Understanding of Technical Artifacts}

%%=============================================================%%
%% GivenName	-> \fnm{Joergen W.}
%% Particle	-> \spfx{van der} -> surname prefix
%% FamilyName	-> \sur{Ploeg}
%% Suffix	-> \sfx{IV}
%% \author*[1,2]{\fnm{Joergen W.} \spfx{van der} \sur{Ploeg} 
%%  \sfx{IV}}\email{iauthor@gmail.com}
%%=============================================================%%

%\author*[1]{\fnm{Anoymous} \sur{Anonymous}}\email{anonymous@anonymous.org}
 \author*[1]{\fnm{Lutz} \sur{Terfloth}}\email{lutz.terfloth@uni-paderborn.de}
 \author[2]{\fnm{Heike} \sur{M. Buhl}}\email{heike.buhl@uni-paderborn.de}
 \author[3]{\fnm{Vivien} \sur{Lohmer}}\email{vivien.lohmer@uni-bielefeld.de}
 \author[2]{\fnm{Michael} \sur{Schaffer}}\email{michael.schaffer@uni-paderborn.de}
 \author[3]{\fnm{Friederike} \sur{Kern}}\email{friederike.kern@uni-bielefeld.de}
 \author[1]{\fnm{Carsten} \sur{Schulte}}\email{carsten.schulte@uni-paderborn.de}
\affil[3]{\orgdiv{Linguistics}, \orgname{Bielefeld University}}
\affil*[1]{\orgdiv{Computing Education Research}, \orgname{Paderborn University}}
\affil[2]{\orgdiv{Psychology}, \orgname{Paderborn University}, \orgaddress{\country{Germany}}}

%%==================================%%
%% Sample for unstructured abstract %%
%%==================================%%

%%================================%%
%% Sample for structured abstract %%
%%================================%%

\abstract{
\textbf{Purpose:} Understanding a technical artifact requires grasping both its internal structure (Architecture) and its purpose and significance (Relevance), as formalized by Dual Nature Theory. This controlled experimental study investigates whether how explainers address these perspectives affects explainees' understanding.

\textbf{Methods:} In a between-subjects experiment, 104 participants received explanations of the board game \textit{Quarto!} from trained confederates in one of three conditions: Architecture-focused (A), Relevance-focused (R), or Integrated (AR). Understanding was assessed on comprehension (knowing that) and enabledness (knowing how).

\textbf{Results:} The A and R conditions produced equivalent understanding despite different explanation content. The AR condition yielded significantly higher enabledness than the focused conditions combined ({\unboldmath$\mathrm{F}(1, 102) = 4.83$, $p = .030$, $\eta^2_p = .045$}), while no differences emerged for comprehension.

\textbf{Conclusion:} Integrating Architecture and Relevance specifically enhances explainees' ability to apply their understanding in practice, suggesting that fostering agency with technical artifacts requires bridging both perspectives. This has implications for technology education and explainable AI design.
}

\keywords{dual nature theory, technical artifacts, explanation, understanding, enabledness, co-construction}

%%\pacs[JEL Classification]{D8, H51}

%%\pacs[MSC Classification]{35A01, 65L10, 65L12, 65L20, 65L70}

\maketitle

\section{Introduction}
When we encounter a technical artifact, be it a smartphone, a sorting algorithm, or a recommendation system, \textit{really} understanding it involves more than knowing what it does. It also requires understanding how it is built and why it was designed the way it was. The Dual Nature Theory formalizes this intuition, positing that technical artifacts possess two essential characteristics: their \textit{Architecture} (how they are built) and their \textit{Relevance} (what they are for) \citep{kroes_engineering_2010,winkelnkemper_ariadne_2024}. \citeauthor{winkelnkemper_ariadne_2024} formalize these as distinct but integrated perspectives on the artifact: the perspective on the Architecture encompasses the physical and formal structures (hardware, algorithms, protocols) while Relevance encompasses functions, societal purposes, discourses, and values. This distinction is not merely techno-philosophical, it has concrete consequences for how explanations of technical artifacts are structured and, as this paper argues, for how well those explanations are understood.

But how are these two perspectives actually addressed when one person explains a technical artifact to another? Fundamental to this question, research on verbal explanation, particularly in face-to-face instructional dialog, has consistently shown that comprehension is shaped by interactive explanatory processes, as speakers adapt explanations to learners' knowledge, monitor understanding, and support the construction of coherent mental representations \citep{chi_icap_2014,foxtree_listening_1999,clark_grounding_1991,graesser_collaborative_1995}. In everyday life, when one person explains a technical artifact to another, given that both parties are motivated, a dynamic process of sense-making unfolds. The explainer selects, frames, and sequences information, while the explainee interprets it against their prior knowledge and interests, asks questions, and communicates confusion \citep{fisher_exploring_2023,rohlfing_explanation_2021}.

Previous work began to characterize this dynamic process empirically by operationalizing the dual nature perspectives as codes for qualitative content coding. A study of board game explanations \citep{terfloth_adding_2023} indicated that explainers naturally draw on both the Architecture and the Relevance perspective when explaining technical artifacts, with an Architecture-first ordering that converges with findings from program comprehension research \citep{pennington_comprehension_1987}. A second study \citep{terfloth_navigating_inpress} further revealed that explainers adapt their explanations to the explainees' interests in Architecture or Relevance while maintaining a broadly Architecture-dominant structure. Together, these findings paint a picture of explanation as a flexible activity in which the two perspectives are navigated somewhat systematically.

Yet an important question has so far remained unanswered: does the particular way in which an explainer addresses the dual nature actually affect how well the explainee understands the artifact? The prior studies observed explanation dynamics as they occurred naturally, but they did not experimentally vary the explanatory approach, making it difficult to draw causal conclusions about explanation effects. The kind and depth of understanding that develops may differ depending on whether explainers emphasize Architecture, foreground Relevance, or explicitly integrate both dimensions. Moreover, the practical stakes of this question are high. If certain patterns of dual nature addressing consistently produce better understanding, such evidence could inform teacher training, curriculum design, and the development of social explainable AI (XAI) systems, where structuring explanations to support not only comprehension but also user agency is an active area of concern \citep{miller_explanation_2019,hoper_data_2023}.

The current study addresses this gap by reversing the observational logic of the previous work. Rather than examining how explainers naturally adapt their explanations, we systematically vary the way they address the dual nature of a technical artifact and examine the causal effect of these variations on understanding outcomes. The study was designed as foundational research: the goal is to \emph{test} theoretically grounded hypotheses under conditions that minimize confounding variables, using an explanandum accessible enough to recruit a broad participant pool without requiring domain-specific prior knowledge. We separate understanding into two dimensions based on \citet{buschmeier_forms_2025}: (1) comprehension (knowing that) and (2) enabledness (knowing how). Only if both dimensions are achieved does a learner not only know what and how an artifact works but can also act with and upon it, a condition \citet{buschmeier_forms_2025} associate with agency. We postulate that the deepest forms of understanding---that is, agency---will be achieved through integrated explanations, which explicitly connect Architecture and Relevance of the explanandum, the two sides of its dual nature. Consequently, we anticipate that various explanatory approaches will generate distinct forms of understanding.

\section{Related Work}
\label{sec:related_work}

\subsection{The Dual Nature of Technical Artifacts}
When we try to understand a technical artifact, we frequently encounter two distinct but interrelated questions: \textit{how does it work} and \textit{what is it for?} The Dual Nature Theory, developed primarily through a research program at Delft University of Technology, formalizes this intuition by positing that technical artifacts possess a dual nature: a physical structure operating according to natural laws and a function rooted in human intentionality \citep{kroes_engineering_2010}. Crucially, neither aspect alone captures the artifact fully. As \citet{kroes_engineering_2010} puts it, the physical conceptualization may account for how the artifact works in terms of physical processes, but without its function, the object loses its status as a technical artifact. Conversely, purely functional descriptions effectively "black box" the physical structure \citep[p.~56]{kroes_engineering_2010}. A holistic understanding would therefore require engaging with both perspectives and, importantly, with the relationship between them, their integration \citep[cf.][]{deridder_reconstructing_2007}.

In accordance with \citet{winkelnkemper_ariadne_2024}, who specifically tailored the dual nature framework to digital artifacts, this paper conceptualizes these two aspects as perspectives on the \textit{Architecture} (how the artifact is constructed and works) and on the \textit{Relevance} (what it is intended for, including purposes, values, and societal significance). This terminology deliberately extends narrow notions of function \citep[see, e.g.,][]{kroes_engineering_2010}. For complex digital artifacts in particular, whose societal impact often extends well beyond their originally intended purposes, a richer notion of Relevance is needed, one that situates artifacts within broader human practices, goals, and values \citep{schulte_framework_2018,winkelnkemper_ariadne_2024}.

\subsection{Understanding Technical Artifacts}
If understanding a technical artifact requires engaging with both its Architecture and its Relevance, a natural follow-up question is: what does such understanding actually look like? Research in computing education has long recognized that understanding technical artifacts involves more than being able to recite how something works. \citet{soloway_learning_1986}, in the context of learning programming, emphasized that novices often learn to construct working programs (mechanisms) while struggling to understand \textit{why} those mechanisms are appropriate. They implement working code without comprehending the goal-plan decompositions \citep[see][]{spohrer_goal_1985} that expert programmers employ as organizing frameworks for design decisions. More recently, \citet{register_learning_2020} found that combining explanations of algorithmic machine learning Architecture with personally meaningful data (expanding the Relevance) led to greater learner agency than explanations with ambiguous data.

This divide has been elaborated in several complementary frameworks. The Block Model \citep{schulte_block_2008} organizes program comprehension into three dimensions (text surface, program execution, and functional goals) across multiple levels of granularity. Expert programmers, on this account, develop flexible mental representations that allow movement between these dimensions, whereas novices tend to become trapped in isolated structural understanding without connecting it to functional purposes \citep{schulte_introduction_2010}. A closely related observation appears in \citeauthor{pennington_comprehension_1987}'s (\citeyear{pennington_comprehension_1987}) work, which differentiates between \textit{program models} (representing Architecture: syntax, control flow, implementation details) and \textit{domain models} (representing Relevance in the form of the problem being solved in a specific context). Pennington found that effective comprehension requires not only the construction of both models but their cross-referencing in a way that connects program parts to domain functions,that connects Architecture to Relevance. 

At a more general level, these perspectives align with cognitive science research on mechanistic versus functional understanding. Drawing on work by \citet{keil_explanation_2006} and \citet{dennett_intentional_2009}, \citet{lombrozo_mechanistic_2019} distinguish between two modes of construal. A \textit{mechanistic mode} orients understanding toward parts, processes, and proximate causal mechanisms; a \textit{functional mode} orients understanding toward functions, goals, and purposes. These are not merely different levels of detail but different strategies for representing dependence relationships: one tracks cause-and-effect chains, the other tracks purpose-and-goal relationships. Lombrozo and Wilkenfeld argue for a \textit{weak differentiation thesis}: mechanistic (Architecture) and functional (Relevance) understanding target different features of the world and afford different inferential advantages, even though the understanding relation itself may be fundamentally the same. Applied to technical artifacts, this implies that understanding the Architecture and understanding the Relevance orients the understander toward genuinely different aspects of how artifacts are built and what they can be used for \citep[cf. discussions in][]{lombrozo_mechanistic_2019}.

The weak differentiation thesis addresses what understanding \textit{targets}, in other words, Architecture and Relevance as different features of the artifact, but it does not address the complementary question of what \textit{form} that understanding takes, whether a person merely knows facts about the artifact or can act competently with it. \citet{buschmeier_forms_2025} address this second dimension by differentiating between \textit{comprehension} (knowing that) and \textit{enabledness} (knowing how to use and adapt a technical artifact). We adopt this distinction for the present study: comprehension denotes the integrated knowledge of both Architecture and Relevance, while enabledness denotes the ability to apply that knowledge in practice.

\subsection{Explanatory Strategies and Understanding Outcomes}

The dual nature framework sketched above provides a foundation for understanding \textit{what} learners need to grasp about a technical artifact. But teaching toward such integrated understanding requires strategic choices about \textit{how} to present information. If learners differ in their interests, prior knowledge, and inferential goals, might explanations strategically adapted to emphasize Architecture or Relevance produce different understanding outcomes?

Prior work on explanation effectiveness provides an important research context. In tutoring settings, \citet{wittwer_can_2010} found that explanations adapted to a learner's current understanding produce stronger outcomes than one-size-fits-all approaches. \citet{nuckles_information_2006} showed more specifically that when tutors possess accurate information about learners' prior knowledge, they make targeted adjustments resulting in more effective communication. What unifies these findings is the idea that understanding requires active construction, not passive reception. Shifting the explanatory emphasis between Architecture and Relevance directly changes what the explanation is aligned with and may therefore change what kind of understanding develops.

This constructive view is reinforced by research on everyday explanation as a social practice. Explanations in face-to-face settings are not unilateral knowledge transfers but co-constructive processes in which explainers and explainees jointly build understanding through dialogue \citep{rohlfing_explanation_2021}. This involves continuous monitoring of mutual understanding, scaffolding of emerging comprehension, and the establishment of common ground \citep{clark_grounding_1991}. For technical artifacts specifically, this implies that how Architecture and Relevance are addressed is not necessarily predetermined but emerges through the interaction, as explainers adapt focus and depth in response to the explainee's evolving interests and understanding. Our previous work underscored this assumption. A first study \citep{terfloth_adding_2023} indicated that explainers naturally draw on both perspectives, with Architecture dominating the early phases before Relevance content increases progressively. A second study \citep{terfloth_navigating_inpress} revealed that explainers detect and adapt to explainees' interests in Architecture or Relevance---a process interpretable as monitoring the explainee's orientation toward one perspective and scaffolding accordingly---while maintaining a broadly Architecture-dominant structure. Together, these findings suggest that the dual nature is navigated somewhat systematically, with both content emphasis and interactional dynamics shaping how the explanation unfolds. 

\subsection{The Bridge Between Architecture and Relevance}

The theoretical case for integrating both perspectives goes beyond simply presenting each one. \citet{kroes_engineering_2010} argues that neither Architecture nor Relevance descriptions alone capture the design of a technical artifact. Instead, genuine understanding requires grasping how the two relate. \citet{deridder_reconstructing_2007} elaborates this point by proposing that explaining technical artifacts is, in a sense, the reverse of the design process: designers bridge from intended function to physical structure, while explainers trace design decisions backward from structure to function. On this account, expert understanding is characterized not by depth in one dimension but by the ability to flexibly move between them.

Empirical evidence supports this claim at several levels. \citet{kallia_search_2023} found that successful programmers employ "Global Backward Explanations" that systematically link code behaviors (Architecture) to problem goals (Relevance), while struggling students focus on implementation details without constructing the crucial interpretive bridges. \citet{miyake_constructive_1986}, studying dyads making sense of sewing machine mechanics, showed that deep understanding emerged through collaborative exploration that iteratively cycled between identifying functions and questioning mechanisms. At the level of text comprehension, \citeauthor{kintsch_role_1991}'s (\citeyear{kintsch_role_1991}) distinction between text-base representations (surface information processing, analogous to Architecture) and situation models (meaning construction through integration with prior knowledge, analogous to Relevance) similarly suggests that understanding requires building interpretive bridges between levels of representation. 

The Architecture-first ordering observed in our previous work \citep{terfloth_adding_2023} adds a temporal dimension to this integration question. Explainers consistently began with structural components and mechanisms before shifting toward Relevance, a pattern that converges with \citeauthor{pennington_comprehension_1987}'s (\citeyear{pennington_comprehension_1987}) finding that the program model is typically constructed before the domain model. At first glance, this ordering stands in tension with cognitive science evidence suggesting that functional information may be psychologically privileged: \citet{mccarthy_right_2023} found that adults prefer explanations presenting functional information before mechanistic information, and \citet{kelemen_function_1999} and \citet{lombrozo_mechanistic_2019} suggest that functional reasoning may be less cognitively demanding than mechanistic reasoning. Yet the tension dissolves on closer inspection. Relevance may be easier to process and may produce a \textit{sense} of understanding, but without a grasp of the Architecture, that sense risks being superficial. The Architecture-first strategy, on this reading, functions as a deliberate epistemic scaffold: it ensures that when Relevance is addressed, the explainee has sufficient structural comprehension to make the connection meaningful. These ordering findings raise a further question that prior work has not addressed: how does the balance between Architecture and Relevance develop over the course of an explanation? If explainers begin with Architecture and shift toward Relevance progressively, as the observational evidence suggests, then the temporal structure of an explanation, and not just its aggregate content, may be relevant to what the explainee understands. 

Together, these lines of evidence suggest that bridging Architecture and Relevance may be crucial for deeper understanding, particularly for forms of understanding that are not just comprehension (knowing that) but also enabledness (knowing how) and through their combination \emph{agency} \citep{buschmeier_forms_2025}. However, while the theoretical arguments and observational evidence are persuasive, direct experimental evidence for an effect of integrated dual-nature explanations on understanding outcomes remains limited. It is this gap the present study addresses.

\section{Research Questions and Hypotheses}
These hypotheses reflect the thesis developed in \autoref{sec:related_work}: Architecture and Relevance represent meaningfully different targets of understanding with different inferential affordances \citep[cf.][]{lombrozo_mechanistic_2019}, and deeper and more flexible understanding emerges when both dimensions are integrated \citep{winkelnkemper_ariadne_2024}, allowing learners to represent and appreciate the dependence relations that connect both. This study addresses whether and how the explainer's approach to addressing the dual nature affects the depth and form of understanding achieved by the explainees:

\medskip\noindent
\textbf{RQ1:} How do different explanatory approaches (Architecture-focused, Relevance-focused, or Integrated) affect the degree and form of understanding achieved by explainees, and which approach best promotes it?

\medskip\noindent
We hypothesize:
\begin{itemize}
    \item \textbf{H1:} Architecture-focused explanations will produce better comprehension but lower enabledness than Relevance-focused explanations.
    \item \textbf{H2:} Relevance-focused explanations will produce better enabledness but lower comprehension than Architecture-focused explanations.
    \item \textbf{H3:} Integrated explanations that explicitly bridge Architecture and Relevance will produce higher levels of both comprehension and enabledness than either focused condition.
\end{itemize}

\medskip\noindent
In addition to the understanding-focused hypotheses above, we address the following exploratory question:

\medskip\noindent
\textbf{RQ2 (exploratory):} How does the distribution of Architecture and Relevance content develop over the course of the explanation across conditions and speakers?

\medskip\noindent
This question is motivated by prior evidence that explainers tend to adopt an Architecture-first strategy \citep{terfloth_adding_2023} and by the theoretical expectation that integration is not merely a matter of overall content balance but of how perspectives are sequenced and connected over time.

\section{Method}
\label{sec:method}
We conducted an experimental study with three between-subjects conditions based on explanatory approach: Architecture-focused (A), Relevance-focused (R), and Integrated (AR). In each session, a confederate explainer trained to deliver condition-specific explanations explained the board game Quarto! to a naive explainee ($N=104$). Quarto! was chosen as an explanandum because it embodies both clear architectural features (pieces, board, rule structure) and meaningful relevance dimensions (strategy, tactics, peculiarities of the game). The game itself was not physically present during the explanation to encourage explicit verbalization. Understanding outcomes were assessed a questionnaire developed and evaluated through an iterative process. The questionnaire is used for measuring comprehension (knowing that) and enabledness (knowing how) \citep{buschmeier_forms_2025}. Additionally, the recorded explanatory dialogues were transcribed and coded for Architecture and Relevance content using qualitative content analysis, serving both as a manipulation check and as data for temporal analysis of how dual nature content unfolded across the interaction.

\subsection{Participants}

\subsubsection{Explainee Participants}
A total of \textit{N} = 104 explainees (EEs) were recruited from [University].\footnote{The total set contained 107 EEs. For the content analysis, one EE was excluded because the participant discontinued the experiment, resulting in 106 coded dyads. For the understanding questionnaire, two participants were excluded based on inattentiveness assessed through the instructional manipulation check (IMC, see \ref{sec:understanding}), and the discontinued participant was likewise absent, yielding the analyzed sample of $N$ = 104. Note that the two IMC-excluded participants remain in the content analysis and sociodemographic data, as the IMC assessed questionnaire attentiveness rather than engagement during the explanation phase. Group sizes for the understanding analyses: $n_A = 36$, $n_R = 35$, $n_{AR} = 33$.} Participants ranged in age from 18 to 60 years (\textit{M} = 23.58, \textit{SD} = 5.05), with 63.21\% identifying as female and 34.91\% as male, with 1.89\% identifying as diverse. All participants were native German speakers with no prior knowledge of the game Quarto!. 

All participants were university students across various degree programs. The largest group studied to become future teachers (Lehramt, $n$ = 46), followed by computer science (Informatik, $n$ = 12) and business-related programs (Wirtschaftswissenschaften, $n$ = 7). Thirty-three percent of participants indicated that their occupation or studies involved explaining ($n$ = 35), while 66.98\% reported no such focus ($n$ = 71).

\subsubsection{Confederate Explainers}

The EX confederates were five student assistants working in the project. They had previously participated as EE confederates in an earlier study \citep{terfloth_navigating_inpress}. As preparation, they completed a standardized 30-minute workshop elaborating the dual nature theory in the context of Quarto!: a 10-minute introduction to the dual nature theory and its application to artifacts, a 10-minute data-session reviewing coded transcripts from a prior naturalistic study of explanation, and a 10-minute categorization quiz where confederates classified utterances into Architecture or Relevance categories, with discussion and resolution of misunderstandings. In pilot sessions, they tested explaining in different conditions, too. Confederates did not memorize their explanations, as their verbal and non-verbal behavior had to remain impromptu and unscripted to maintain naturalness. The setting and confederate training effectiveness were qualitatively assured across five pilot studies. All confederates explained in all conditions to control for confederate-specific effects. Confederate fidelity to condition instructions was additionally assessed via content analysis of post-session recordings (see \autoref{sec:data_analysis}), confirming that manipulated conditions aligned with their intended emphasis on Architecture, Relevance, or integration of both.

\subsection{Materials}

\subsubsection{Explanandum}
Following prior research  \citep{terfloth_adding_2023,terfloth_navigating_inpress}, we deliberately chose Quarto! because it exemplifies a technical artifact with distinct and observable Architecture (physical components, governing rules, board layout) and Relevance (tactical approaches, aesthetic choices). Importantly, Quarto! enabled us to examine core explanatory mechanisms in relative isolation, minimizing confounding factors that complicate analysis of more sophisticated technological systems—such as explainer uncertainty or limited artifact transparency. Quarto! also allows comparison of results with previous studies.

Quarto! features 16 distinct pieces, each defined by four binary attributes (tall/short, light/dark, square/round, solid/hollow). Players alternate placing pieces on a 4×4 grid, aiming to complete a line of four pieces sharing at least one common attribute. A distinctive rule allows players to determine which piece their opponent must place, introducing strategic depth. Quarto! serves as an ideal explanandum because it embodies both clear architectural features (pieces, board, rule structure) and meaningful relevance dimensions (strategy, tactics, victory conditions, potential mistakes). Despite being straightforward to learn, mastering it proves challenging.

\subsubsection{Explanation Conditions}

To examine how deliberate foci in explanatory emphasis affect understanding, these experimental conditions deliberately exaggerated the focus on one side of the dual nature to create experimentally separable treatments:

\begin{itemize}
\item \textit{Architecture-focused (A):} Predominantly emphasized game mechanics, rules, piece properties, and structural relationships. Confederate EXs were instructed to allocate the majority of their explanatory effort to Architecture content, while including only minimal-but-essential Relevance content necessary to ensure basic coherence and conversational naturalness.

\item \textit{Relevance-focused (R):} Predominantly emphasized strategies, goals, tactical considerations, and gameplay dynamics. Confederate EXs were instructed to allocate the majority of their explanatory effort to Relevance content, while including only minimal-but-essential Architecture content necessary to ensure basic coherence and conversational naturalness.

\item \textit{Integrated (AR):} Balanced presentation of both Architecture and Relevance with explicit bridging statements. Confederate EXs were instructed to address both Architecture and Relevance approximately equally, with deliberate emphasis on how design choices (Architecture) enable or constrain strategic possibilities (Relevance).
\end{itemize}

\subsection{Procedure}

Upon arrival, participants (EEs) were welcomed, guided to the explanation room, and asked to read and sign an informed consent form. They then provided payment information for reimbursement (10 € per hour or via study credits applicable to Psychology students) and completed a sociodemographic questionnaire. Before the explanation phase commenced, participants were instructed to set their phones to flight mode, and, if necessary, to dispose of any chewing gum. 

The EX confederate was then guided into the explanation room, where both participants took their designated seats. At this point, EEs believed EXs to be naive participants as well. Camera recordings were initiated with a clap for synchronization during post-processing. The EX received a standardized instruction: \textit{"Explain the board game Quarto! in such a way that the other person could win."} The EE received the instruction to take part in the explanation actively. The explanation ended naturally when the EE had no further questions and the EX considered it complete, ensuring no predetermined duration or time pressure. The board game Quarto! was \textit{not} present during the explanation. We excluded the physical game for several reasons. We anticipated that having the game present would discourage explainers from explicitly articulating Architecture details (such as describing the piece shapes). By withholding the game, EXs must consciously select which elements warrant explanation, revealing the reasoning strategies they employ, for instance, how they abstract or generalize specific features. On the theoretical premise that gestures reflect underlying cognitive activity \citep{mcneill_hand_1992,streeck_gesturecraft_2009}, we also aimed for a dataset suitable for gesture analysis in future work. The table was therefore kept clear of physical objects.

Following the explanation, the EE completed a post-explanation questionnaire containing an item to assess naturalness of the explanation (rated 1-7). Participants were then offered the opportunity to play two rounds of Quarto! together if they wished. A standardized verbal debriefing followed, in which the participants (EEs) were informed of the confederate design and asked to maintain confidentiality. 

Explanation duration varied considerably across dyads. Overall, dialogues lasted \textit{M} = 410.1 seconds (\textit{SD} = 176.5; 95\% CI [376.1, 444.1]), approximately 6.8 minutes on average. Condition-level descriptives revealed substantial variation: the Architecture condition produced the shortest explanations (\textit{M} = 265.7s, \textit{SD} = 89.9), followed by the Integrated condition (\textit{M} = 454.6s, \textit{SD} = 181.8), and the Relevance condition (\textit{M} = 514.1s, \textit{SD} = 139.8). A one-way ANOVA confirmed a significant condition effect on dialog duration, $F(2, 103) = 29.81$, $p < .001$. Post-hoc comparisons (Tukey HSD) indicated that the Architecture condition produced significantly shorter explanations than both the Relevance condition ($p < .001$) and the Integrated condition ($p < .001$), whereas the Relevance and Integrated conditions did not differ ($p = .190$). Explanation length is therefore a potential confound when comparing understanding outcomes across conditions; we return to this in the Discussion.

\subsection{Measures}
\label{sec:understanding}

We operationalized understanding through two forms, following the distinction introduced in the Related Work section: \textit{comprehension} (knowing that) as knowledge of both architectural and relevance aspects and their integration, and \textit{enabledness} (knowing how) as the ability to apply that understanding in practice \citep[cf.][]{buschmeier_forms_2025}. Comprehension comprises two sub-dimensions: Architecture comprehension (knowing that regarding game mechanics, rules, components, and structural relationships) and Relevance comprehension (knowing that regarding strategies, goals, and functional significance of design choices). Enabledness captures the ability to apply that comprehension in practical game scenarios. Understanding outcomes were operationalized as three mean scores: a \textit{comprehension score} (mean of correct responses on Architecture and Relevance items combined), an \textit{enabledness score} (mean of correct responses on scenario-based Enabledness items), and a \textit{total score} (mean of both). Mean scores were used rather than sum scores to facilitate interpretability on the original item metric (0--1 scale, representing proportion of correctly answered items). 

To measure understanding along these dimensions, we developed and validated a questionnaire through an iterative process involving interview-based item generation, two rounds of exploratory factor analysis on independent samples (\textit{N} = 157, \textit{N} = 134), and a final evaluation. The full development and validation process is documented in an OSF repository\footnote{\url{https://osf.io/w39dc/overview?view_only=517962d50d314c529b6d0aafe20cf650}}. The final instrument comprised 34 items across three subscales: 13 Architecture items, 12 Relevance items, and 6 Enabledness items, plus 1 Instructional Manipulation Check item assessing inattentiveness and 2 items for participant identification across questionnaires. Architecture and Relevance items use a binary Agree/Disagree format; Enabledness items present visual game scenarios requiring participants to select the correct answer from multiple-choice options (between 3 and 8 options).

Item assignments follow the conceptual structure established during scale development. However, in the present study the Architecture and Relevance subscales did not function as separable dimensions: Architecture and Relevance items loaded together on shared factors in both exploratory factor analyses during development, and confirmatory factor analysis in the present sample (\textit{N} = 104) did not yield admissible solutions, likely due to the binary item format and ceiling effects produced by effective explanations. Internal consistency for the combined comprehension subscale was $\alpha$ = .65, and $\alpha$ = .49 for the Enabledness subscale. The correlation between Architecture and Relevance comprehension subscores was moderate ($r = .35$, $p < .001$), consistent with the EFA pattern of partial overlap between the two dimensions in this domain. We therefore do not treat Architecture and Relevance comprehension as separable outcomes in the present analyses. Instead, we report a single aggregated comprehension score (Architecture + Relevance items combined) alongside the enabledness score.

\subsection{Data Analysis}
\label{sec:data_analysis}

\subsubsection{Content Coding}

Recordings of the explanatory dialogues were automatically transcribed using Whisper, with manual verification by student assistants following established transcription conventions \citep[rules 1, 3, 4, 14]{kuckartz_transcribing_2019}. All utterances were segmented into intonation units \citep[cf.][]{selting_gesprachsanalytisches_2009}, prosodic units of spoken language that enhanced coding reliability in previous applications. This resulted in each transcript line corresponding to a single intonational unit. More detail on the methodological approach and rationale for using intonation phrases can be found in \citep{terfloth_transcription_2025}.

We employed qualitative content analysis to examine explanation content, drawing on a coding manual developed and validated in prior work \citep{terfloth_adding_2023,terfloth_navigating_inpress}. Two dimensions structured the coding scheme. First, we classified utterances into two content categories: those addressing the Architecture of Quarto! and those addressing its Relevance. A detailed coding manual specifies category boundaries and edge cases\footnote{\url{https://osf.io/29pzu/overview?view_only=0b4e0a7546c0435b8b1ded9c8ecd43f7}}. Second, we identified the speaker for each utterance (EXs or EEs). Two student assistants coded the data using ELAN \citep{wittenburg_elan_2006}\footnote{ELAN is an open-source multimedia annotation tool provided by the Language Archive at Max Planck Institute for Psycholinguistics (Nijmegen, the Netherlands) that enables precise temporal alignment of annotations with audio and video recordings. \url{https://archive.mpi.nl/tla/elan}}. In two earlier studies, the coding manual's adequacy was confirmed ($\kappa=.81$) in settings with the same coders and almost equal design (i.e., one uncontrolled explanation setting, one where EEs were confederates); therefore intercoder reliability was not assessed in the present study. All coders were blind to condition assignment throughout annotation. Across the 115,155 coded segments in the transcripts, the distribution was as follows: 43.2\% were coded as Architecture, 36.3\% as Relevance, 20.3\% received no code (e.g., conversational filler), and 0.3\% received both codes. Representative examples are: "And these pieces have four properties: they are either tall or small, hollow or solid, dark or light, round or square" (Architecture), "so what makes this game interesting is that, if you placed one piece, then you decide which piece your opponent places next" (Relevance), "hey I am name" (none) and "Quarto! is similar to connect four" (both).

\subsubsection{Temporal Analysis of Explanation Content}

To compare how explanations distributed content across Architecture and Relevance despite varying lengths, we developed a temporal normalization approach. We identified each transcript's substantive boundaries, specifically, the first utterance mentioning Quarto and the last one, and rescaled these intervals to a 0--1 timeline. This normalization enabled us to represent each coded segment as a proportion of the explanation's overall temporal span (e.g., an Architecture code occurring early might occupy the 0--0.03 range). By treating coding categories as continuous distributions across this normalized time axis, we could characterize how explanations unfolded and directly compare patterning across different speakers and experimental conditions. We further validated whether the experimental conditions successfully shaped the balance of Architecture and Relevance content by examining the proportion of each category produced by confederates in each condition (see \autoref{sec:results}). To capture phase-dependent dynamics, we divided each normalized timeline into three equal segments (first, second, and third third), representing the opening, middle, and closing phases of the explanation. This division is coarse enough to yield robust per-phase estimates even for shorter dialogues, while fine enough to capture the opening-to-closing progression that prior work identified as characteristic of naturalistic explanations \citep{terfloth_adding_2023}. For each dyad, we computed the proportion of Architecture and Relevance content within each third, separately for EX and EE utterances and for both.

\subsubsection{Understanding Outcomes}
We first tested for differences across the three conditions (A, R, AR) using one-way ANOVAs and non-parametric Mann--Whitney $U$ tests reported alongside as a robustness check. Hypothesis 3 predicts a contrast between single-focus explanations and integrated explanations, we additionally conducted planned comparisons collapsing the A and R conditions into a single "focused" group and comparing it to the AR group. Effect sizes are reported as partial eta-squared ($\eta^2_p$) for ANOVAs and as Hedges' $g$ and rank-biserial correlations for pairwise comparisons.

\subsubsection{Statistical Approach}
\label{sec:stat_approach}
 
The primary understanding outcomes are derived from binary and multiple-choice items. Because such scores can violate distributional assumptions of parametric tests, we adopt the following strategy throughout the understanding analyses: parametric tests (one-way ANOVAs, Welch's $t$-tests) serve as the primary analyses, supplemented by non-parametric equivalents (Kruskal--Wallis $H$, Mann--Whitney $U$) as robustness checks. Where parametric and non-parametric results converge, we report only the parametric statistics and note convergence; where they diverge, we report both and interpret conservatively. For the treatment check, which relies on content proportions rather than participant-level outcome scores, parametric and non-parametric tests consistently converged (all $p$s $< .001$); we therefore report only parametric results there. Continuous variables (e.g., dialog duration, naturalness ratings) are analyzed with parametric tests only.

\section{Results}
\label{sec:results}

\subsection{Confederate Treatment Check}
\label{sec:treatment}
\begin{figure}
    \centering
    \includegraphics[width=1\linewidth]{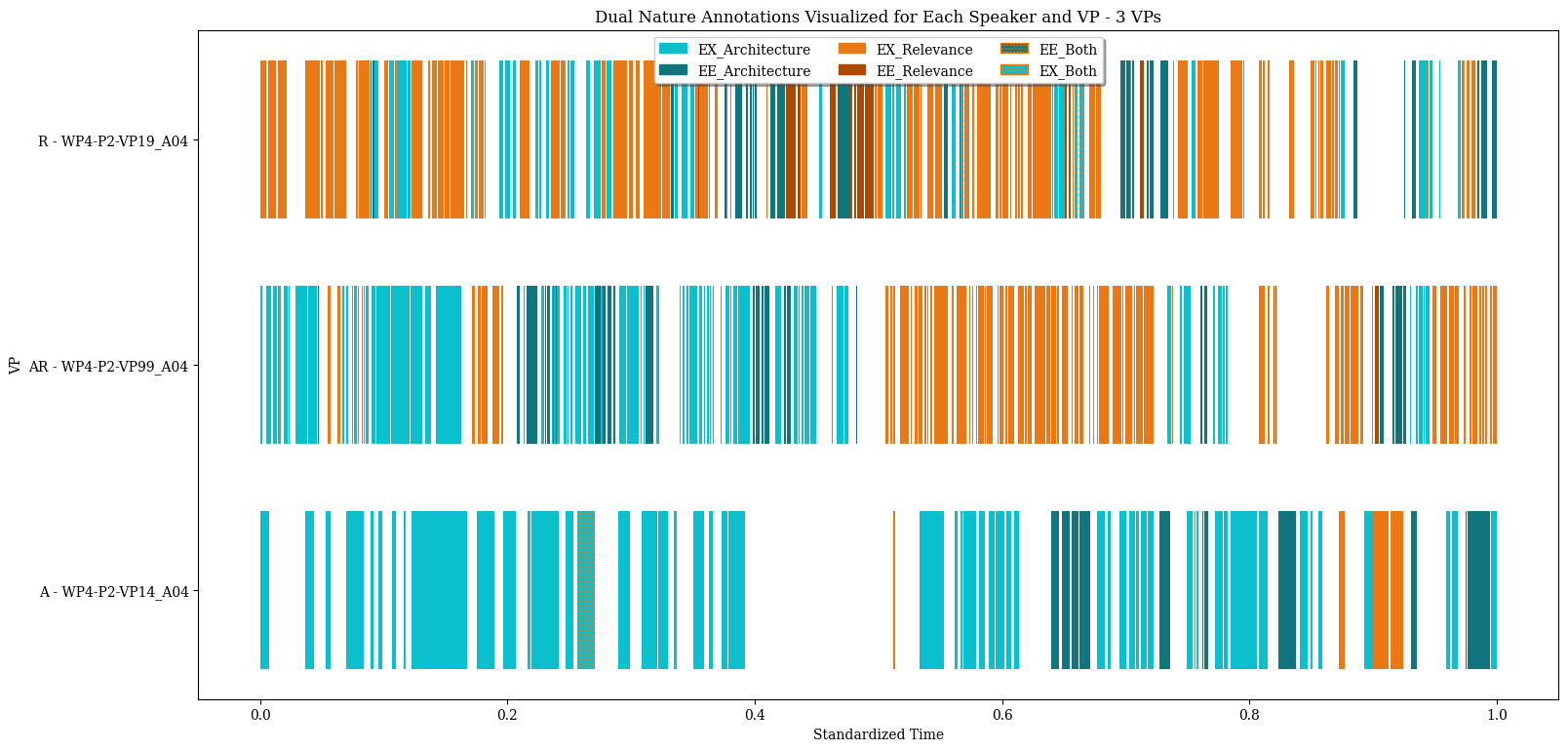}
    \caption{Illustrative examples of dual nature annotation timelines for one representative dyad per condition. Each bar represents the normalized dialog timeline (0--1), with segments colored by coding category: Architecture (blue), Relevance (orange), and uncoded content (white), with different shading per speaker. Dyads were selected as the closest to their condition mean in Euclidean distance on Architecture and Relevance proportions (A: arch = 92.4\%, rel = 7.6\%; R: arch = 36.5\%, rel = 63.5\%; AR: arch = 55.5\%, rel = 44.5\%).}
    \label{fig:illustrative_comparison}
\end{figure}

Content analysis of coded EX utterances in the explanation transcripts confirmed successful manipulation across conditions ($N$ = 95,590 coded EX utterances across 106 dyads). All pairwise comparisons between conditions were highly significant for both Architecture and Relevance proportions (all $ps < .001$ by Welch's $t$-tests and Mann--Whitney $U$ tests; see Table~\ref{tab:manipulation}).

\begin{wraptable}[12]{r}{0.55\textwidth}
    \centering
    \label{tab:manipulation}
    \begin{tabular}{lccc}
    \toprule
    \multicolumn{4}{c}{Condition} \\
    \cmidrule(lr){1-4}
    Content  & A & R & AR \\
    \midrule
    Architecture & 71.7\% & 26.0\% & 44.4\% \\
    Relevance & 5.9\% & 57.1\% & 38.7\% \\
    Both & 0.2\% & 0.3\% & 0.2\% \\
    None (uncoded) & 22.1\% & 16.6\% & 16.7\% \\
    \bottomrule
    \end{tabular}
    \vspace{.5em}
    \caption{Observed proportions of Architecture and Relevance content by condition, based on coded EX utterances. ``None'' denotes utterances classified as neither category (e.g., back\-channels, fillers, smalltalk).}
\end{wraptable}
Participants rated the interactions as similarly natural across all three conditions. Ratings were high in the A-condition ($M = 5.06$, $SD = 1.74$, $95\% \text{ CI } [4.49, 5.62]$), the AR-condition ($M = 5.40$, $SD = 1.42$, $95\% \text{ CI } [4.93, 5.87]$), and the R-condition ($M = 4.97$, $SD = 1.52$, $95\% \text{ CI } [4.47, 5.48]$). A one-way ANOVA revealed no significant differences in naturalness ratings across conditions ($F(2, 103) = 0.74$, $p = .482$). Pairwise comparisons confirmed no significant differences (all $p > .22$, all $|d| < .30$).

To verify that the experimental conditions were maintained consistently throughout the explanations, we examined how EX Architecture proportions varied across conditions and over the course of the dialogue (divided into thirds). A mixed ANOVA confirmed that the conditions differed strongly in their Architecture content ($\eta^2_p = .789$, a large effect), and that this difference shifted systematically across dialogue thirds ($\eta^2_p = .333$), meaning that the three conditions did not simply differ in overall amount but also in how their content emphasis developed over time. These temporal dynamics are described in more detail Section~\ref{sec:temporal}.

\subsection{Temporal Patterns in Explanation Content}
\label{sec:temporal}

\begin{figure}
    \centering
    \includegraphics[width=1\linewidth]{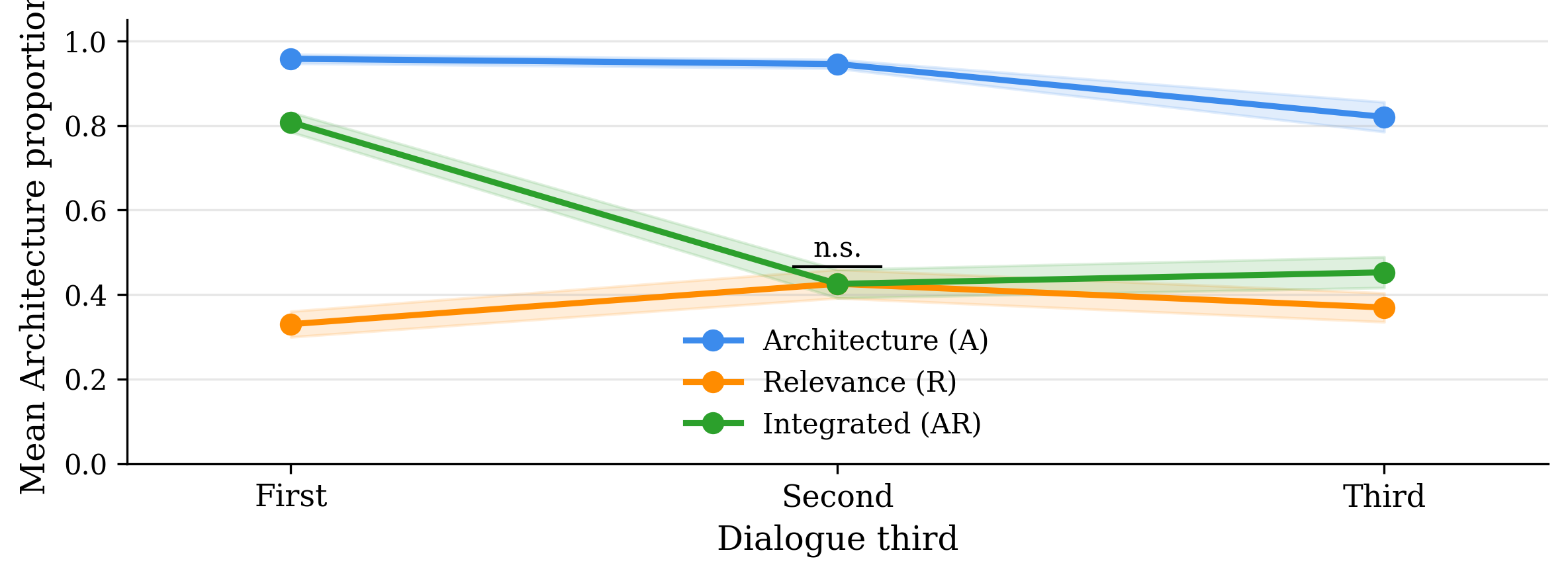}
    \caption{Mean Architecture proportion across dialogue thirds by condition, averaged over both EX and EE utterances ($\pm$\,SEM). The A and AR conditions open with Architecture-dominant content, while the R condition is Relevance-oriented from the outset ($M_\text{arch} = .33$ in the first third). The AR condition shifts toward more Relevance-oriented content across the dialogue, converging with R in the second third ($p=.998$) and remaining statistically indistinguishable from R in the third ($p = .100$). The A condition remains Architecture-dominant throughout, declining gradually from $M = .96$ to $M = .82$.}
    \label{fig:temporal_thirds}
\end{figure}

Having established that the experimental conditions produced distinct content profiles (Section~\ref{sec:treatment}), we turn to RQ2: how did the balance between Architecture and Relevance develop over the course of the explanation? We first report patterns across all utterances before examining EX and EE contributions separately. Analysis of how content was distributed across dialogue thirds revealed condition-specific temporal patterns (see Table~\ref{tab:temporal} and Figure~\ref{fig:temporal_thirds}). Overall, all three conditions differed significantly from one another (A vs.\ AR: $t(69) = 16.22$, $p < .001$; A vs.\ R: $t(69) = 18.44$, $p < .001$; AR vs.\ R: $t(68) = 5.24$, $p < .001$). Within this overall difference, however, a temporal pattern emerged: Architecture content was dominant in the opening phase for the A and AR conditions, while the R condition was Relevance-oriented from the outset ($M_\text{arch} = .33$ in the first third). AR and R converged in the second third ($p = .998$) and remained statistically indistinguishable in the third ($p = .100$).

\begin{wraptable}[24]{r}{0.5\textwidth}
\centering
\caption{Significant Pairwise Differences in EX Content Proportions by Dialogue Third}
\label{tab:temporal}
\begin{tabular}{llcc}
\toprule
Third & Comparison & $t$ & $p$ \\
\midrule
First & A vs.\ AR & 6.29 & $<.001$ \\
First & A vs.\ R & 21.56 & $<.001$ \\
First & AR vs.\ R & 13.25 & $<.001$ \\
Second & A vs.\ AR & 14.94 & $<.001$ \\
Second & A vs.\ R & 15.63 & $<.001$ \\
Third & A vs.\ AR & 7.66 & $<.001$ \\
Third & A vs.\ R & 10.64 & $<.001$ \\
Third & AR vs.\ R & 2.80 & .007 \\
\bottomrule
\end{tabular}
\begin{flushleft}
\textit{Note.} Welch's $t$-tests on EX Architecture proportions. Non-significant comparisons omitted. AR vs.\ R was non-significant in the second third ($t = 1.22$, $p = .23$). The third-third divergence between AR and R ($p < .01$) is specific to the explainer level; when all speakers are combined, AR and R remain statistically indistinguishable in the third third ($t(68) = 1.67$, $p = .10$).
\end{flushleft}
\end{wraptable}
At the EX level, the first third showed the expected ordering: A highest ($M = .96$), followed by AR ($M = .81$) and R ($M = .31$), with all pairwise comparisons significant (all $p$s $< .001$; see Table~\ref{tab:temporal}). By the second third, AR and R EXs converged ($t = 1.22$, $p = .23$): AR EXs shifted toward Relevance content, producing a profile statistically indistinguishable from R EXs. In the third third, a modest divergence re-emerged ($t = 2.80$, $p = .007$). The A condition remained Architecture-dominant throughout, though with a gradual decline from .96 to .82, suggesting that even under explicit instructions to maintain an Architecture focus, some drift toward Relevance occurred.

EE content patterns complemented the EX findings but are reported more briefly here. A corresponding ANOVA on EE Architecture proportions revealed a significant Condition effect ($\eta^2_p = .254$). The central pattern was that A-condition EEs contributed distinctly more Architecture content than either AR-condition EEs ($t(35.81) = 5.38$, $p < .001$, $g = 1.51$) or R-condition EEs ($t(36.60) = 4.64$, $p < .001$, $g = 1.24$), while AR and R EEs were statistically indistinguishable ($t(47.80) = -0.26$, $p = .793$, $g = -0.07$). This divergence was not present in the first third ($p = .433$) but emerged progressively through the second and third thirds (both $p < .001$), suggesting a cumulative effect of sustained exposure to Relevance content rather than an immediate reorientation.

With the content structure of the explanations established, both in aggregate and across phases, we now turn to the central question: did these different explanatory approaches produce different understanding outcomes?

\subsection{Understanding Outcomes}

\subsubsection{Three-Condition Comparison}

Shapiro--Wilk tests indicated departures from normality for enabledness in all conditions and comprehension in A and AR (all $p \leq .002$), consistent with ceiling effects and motivating the dual-testing strategy described in Section~\ref{sec:stat_approach}.

One-way ANOVAs comparing the three conditions (A, R, AR) revealed no significant differences for comprehension, $F(2, 101) = 0.70$, $p = .500$, $\eta^2_p = .014$ ($N = 104$; two participants excluded based on IMC, see Section~\ref{sec:understanding}), or for enabledness, $F(2, 101) = 2.50$, $p = .087$, $\eta^2_p = .047$. Non-parametric Kruskal--Wallis tests converged: comprehension, $H(2) = 1.78$, $p = .411$; enabledness, $H(2) = 5.27$, $p = .072$.

Pairwise comparisons confirmed this pattern. For comprehension, no pair differed significantly (all $p > .15$ by Mann--Whitney $U$; all $p_{\text{Holm}} > .76$ by Welch's $t$). For enabledness, the only noteworthy comparison was AR vs.\ R, which reached significance in the uncorrected Mann--Whitney test ($U = 759.0$, $p = .020$) but was marginal after Holm correction ($p_{\text{Holm}} = .064$). No other enabledness pair approached significance (A--AR: $U = 479.5$, $p = .156$; A--R: $U = 694.0$, $p = .446$).

\subsubsection{Focused vs.\ Integrated Comparison}

To test H3, given that A and R conditions produced equivalent outcomes and that the theoretical framework predicts a contrast between single-focus and integrated explanations, we conducted a comparison collapsing A and R into a single ``focused'' group (A or R) versus the AR group. Results are summarized in Table~\ref{tab:focused_vs_integrated}.
 
\begin{table}[h]
\centering
\caption{Comparison of Focused (A or R) vs.\ Integrated (AR) Conditions}
\label{tab:focused_vs_integrated}
\begin{tabular}{lcccccc}
\toprule
Measure & $F(1,102)$ & $p$ & $\eta^2_p$ & $U$ & $p_U$ & Hedges' $g$ \\
\midrule
Comprehension & 0.88 & .349 & .009 & 1000.0 & .228 & $-0.20$ \\
Enabledness   & 4.83 & .030 & .045 & 875.5  & .032 & $-0.46$ \\
\bottomrule
\end{tabular}
\vspace{0.5em}
\begin{flushleft}
\textit{Note.} $F$ and $\eta^2_p$ from one-way ANOVA; $U$ and $p_U$ from Mann--Whitney $U$ test; Hedges' $g$ from Tukey HSD pairwise comparison. Negative Hedges' $g$ values indicate higher mean scores in the AR group. Both parametric and non-parametric tests converge for comprehension (both non-significant) and for enabledness (both significant at $\alpha = .05$).
\end{flushleft}
\end{table}
 
Participants in the AR condition achieved significantly higher enabledness scores than participants in the focused conditions. This effect was consistent across parametric and non-parametric tests: $F(1, 102) = 4.83$, $p = .030$, $\eta^2_p = .045$; $U = 875.5$, $p = .032$; Hedges' $g = -0.46$. No significant difference was observed for comprehension by either test: $F(1, 102) = 0.88$, $p = .349$; $U = 1000.0$, $p = .228$.
 
We note that for total score, parametric and non-parametric results diverged: the ANOVA was non-significant, $F(1, 102) = 3.25$, $p = .075$, $\eta^2_p = .031$, while the Mann--Whitney $U$ test reached significance, $U = 887.5$, $p = .046$. Following our pre-stated strategy of interpreting conservatively where tests diverge, we do not claim a significant total score difference but note that the marginal parametric result ($p = .075$) is consistent with the interpretation that the enabledness signal is diluted when combined with 25 comprehension items that show no condition difference.

\subsubsection{Exploratory Analyses}
\label{sec:switch_analysis}

As an exploratory complement to the temporal analysis, we computed the rate of Architecture--Relevance content transitions per dyad (i.e., the proportion of consecutive utterance pairs in which the content category switched) and examined its correlation with understanding outcomes. The overarching idea was that more frequently bridging the perspectives might be a strategy that generally outperforms other approaches. But no significant associations were observed (all Spearman $r < .17$, all $p > .12$).

As a further exploratory complement, we examined whether the degree of Architecture emphasis in the opening phase of AR explanations predicted understanding outcomes. Within the AR condition ($n = 33$), the EX Architecture proportion in the first dialogue third was not significantly correlated with enabledness ($r_s = .061$, $p = .738$), comprehension ($r_s = -.108$, $p = .550$), or total score ($r_s = -.084$, $p = .642$). This null result complements the switch-rate finding: within integrated explanations, neither transition frequency nor opening emphasis predicted understanding, suggesting that the benefit of integration is not reducible to a particular sequencing strategy. However, the small subsample limits the detectable effect size, and this finding should be treated as hypothesis-generating.

\section{Discussion}

\subsection{Summary of Main Findings}
This study investigated whether the explainer's approach to addressing the dual nature of technical artifacts---emphasizing Architecture, emphasizing Relevance, or explicitly integrating both---produces different forms of understanding. The results partially support our hypotheses but reveal an unexpected pattern that refines the theoretical picture.

Contrary to H1 and H2, the Architecture-focused and Relevance-focused conditions did not produce the predicted differential profiles. H1 predicted that Architecture-focused explanations would yield better comprehension but limited enabledness, while H2 predicted that Relevance-focused explanations would yield better enabledness but limited comprehension. Instead, the A and R conditions produced virtually identical outcomes on all measures ($p = .786$), despite the treatment check confirming radically different explanation content.

H3 predicted that integrated explanations would produce the highest levels of comprehension and enabledness. This hypothesis was partially supported: the AR condition produced significantly higher enabledness scores than the focused conditions ($F(1, 102) = 4.83$, $p = .030$). The predicted advantage on comprehension did not materialize ($F(1, 102) = 0.88$, $p = .349$). The integration advantage was thus specific to the enabledness dimension of understanding.

\subsection{Evaluation of Hypotheses}

\textbf{H1 and H2: Not Supported.} The most striking finding is the complete equivalence of the A and R conditions. We had expected that each focused condition would produce a characteristic understanding profile: Architecture explanations fostering comprehension, Relevance explanations fostering enabledness. The absence of any differential effect has several possible explanations. First, the comprehension items in the current instrument do not differentiate finely enough between architecture-specific and relevance-specific knowledge, as they are aggregated into a single comprehension score, which may have masked condition-specific advantages. Second, within the constrained domain of a board game, even single-focus explanations may convey enough implicit information from the other dimension to produce comparable outcomes, particularly given that even the R condition included 29.7\% Architecture content. This suggests that some minimal Architecture content may function as a necessary scaffold even in Relevance-focused explanations, consistent with prior work documenting that Architecture serves as a foundation upon which Relevance content is built \citep{terfloth_adding_2023}. Third, the ceiling effects observed in several comprehension items may have compressed variance, making condition differences undetectable.

\medskip
\noindent
\textbf{H3: Partially Supported.} The prediction that integrated explanations would produce the highest enabledness was confirmed. The AR condition produced significantly better enabledness than either focused condition. However, the predicted advantage on comprehension was not found. This partial support suggests that integration of Architecture and Relevance is specifically relevant for the enabledness dimension, the capacity to apply knowledge in practical scenarios, rather than for comprehension per se. The marginal effect on total score ($p = .075$ parametric, $p = .046$ non-parametric) is consistent with this interpretation: the enabledness signal is substantially diluted when combined with 25 comprehension items that show no condition difference, producing a weaker aggregate effect.

\subsection{Theoretical Implications}

\textbf{Why Integration Enhances Enabledness but Not Comprehension.} The selective effect on enabledness, while not predicted in this specific form, is interpretable within the dual nature framework. Comprehension may be achievable through exposure to either side of the duality. Whether an EX emphasizes Architecture or Relevance, the factual content conveyed may be sufficient for EEs to answer knowledge-based questions. In contrast, enabledness, the capacity to actually use the artifact competently in novel situations, plausibly requires connecting knowledge about Architecture and Relevance. The scenario-based enabledness items required participants to make tactical decisions in concrete game situations, which demands integrating knowledge about \textit{how the game works} (Architecture) with knowledge about \textit{what matters strategically} (Relevance). Only the AR condition provided explicit bridges between these perspectives. This post-hoc interpretation aligns with \citet{schulte_framework_2018} and \citet{deridder_reconstructing_2007} who argue that genuine understanding of technical artifacts requires bridging both sides of the duality, while refining this claim: such bridging appears to be specifically necessary for the enabledness dimension rather than for comprehension.

\medskip
\noindent
\textbf{The Unexpected Equivalence of Focused Conditions.} The A $\approx$ R equivalence challenges the intuitive assumption that explanation content directly determines what is learned. Despite radically different content profiles (71.7\% vs.\ 26.0\% Architecture; 5.9\% vs.\ 57.1\% Relevance), participants achieved virtually identical understanding outcomes. Several factors may explain this. First, Architecture and Relevance may be sufficiently interconnected that EEs spontaneously infer one from the other---and even R-condition explanations still contained 26.0\% Architecture content, suggesting some minimal structural scaffolding persists even under Relevance-focused instruction. Second, the comprehension measure aggregates Architecture and Relevance items into a single score (motivated by the EFA-based factor structure), which may mask condition-specific advantages at the subscale level. Third, ceiling effects on several items may have compressed variance, making differences undetectable. A more complex explanandum would likely reduce both the ceiling effects and the cross-inferability between perspectives, potentially revealing the differential profiles that H1 and H2 predicted.

\medskip
\noindent
\textbf{Reconciling Natural and Optimal Patterns.} 
\citep{terfloth_adding_2023} revealed that EXs naturally employ Architecture-first strategies, and the present data confirm that even experimentally controlled explanations tend to open with Architecture-dominant content across all conditions. The present findings do not challenge the utility of this natural pattern for comprehension, as the A condition achieved comprehension on par with all other conditions. However, the data suggest that the Architecture-first strategy, if not supplemented by explicit integration with Relevance, is insufficient for promoting enabledness. To promote agency, understood here as the condition in which both comprehension and enabledness are achieved \citep[cf.][]{buschmeier_forms_2025}, EXs need to progress beyond Architecture toward explicit integration with Relevance. The exploratory correlation analysis within the AR conditions first third reinforces this interpretation. The degree of Architecture emphasis in the opening phase was unrelated to any understanding outcome ($r_s = .061$ for enabledness, $r_s = -.108$ for comprehension, both $p > .55$; $n = 33$). Combined with the null result for content switch rates, this suggests that the benefit of integrated explanations is not attributable to a particular sequencing strategy---neither an Architecture-first opening nor frequent perspective switching predicted better outcomes. What appears to matter is instead the presence of explicit integration between Architecture and Relevance in a qualitative, meaningful sense, regardless of how the explanation is temporally organized---though the small subsample ($n = 33$) limits detection to large effects, and moderate associations cannot be ruled out. In the present study, producing integrated explanations required targeted confederate training, including explicit instruction on the dual nature framework and practice connecting Architecture to Relevance.\citet{frederik_teaching_2011} provide independent evidence that this capacity cannot be taken for granted: in their study, both experienced and inexperienced teachers had difficulty even differentiating structural properties from functional tasks, let alone systematically connecting them. Whether untrained explainers spontaneously produce such bridging remains an open question, but we believe it is not intuitively achieved, implying without explicit curricular emphasis on connecting Architecture and Relevance, it might not become part of teachers' explanatory repertoire.

\subsection{Exploratory Findings: Temporal Dynamics and Co-Constructive Patterns}

The temporal and speaker-level patterns reported under RQ2 were not predicted by the hypotheses and should be interpreted as exploratory. They do, however, offer suggestive evidence for how dual-nature content shapes the co-constructive dynamic between explainer and explainee.

\medskip
\noindent
\textbf{Architecture Drift in the A Condition.} The Condition $\times$ Third interaction ($\eta^2_p = .333$) revealed that even EXs instructed to maintain an Architecture focus showed a gradual decline in Architecture proportion from .96 to .82 across dialogue thirds. This suggests that pure Architecture explanations may be inherently unstable: as enough structural foundation is laid, the conversational context pulls explanations toward strategic and functional content. The AR condition, rather than fighting this natural pull, channels it deliberately through bridging statements that connect Architecture to Relevance. That the AR condition produced the only significant understanding advantage (on enabledness) is consistent with the interpretation that deliberate integration outperforms both unchecked drift and forced single-perspective maintenance.

\medskip
\noindent
\textbf{The Asymmetric Influence of Relevance on Explainee Engagement.} EEs contributed predominantly Architecture-oriented content across all conditions (49.1\% Architecture vs.\ 17.9\% Relevance overall), suggesting that when learners spontaneously engage with a technical artifact, they gravitate toward structural features. However, this overall Architecture dominance conceals a significant condition effect. A-condition EEs contributed distinctly more Architecture content than EEs in either the AR or R conditions (both $p < .001$, $g > 1.2$), while AR and R EEs were statistically indistinguishable ($g = -0.07$). This pattern suggests a ratchet-like dynamic: Architecture content alone produces Architecture-focused EEs, but the introduction of any sustained Relevance content, whether in the R or AR condition, shifts EEs toward a more balanced engagement pattern. The shift was not immediate---A and AR EEs did not differ in the first third ($p = .433$)---but emerged cumulatively through the second and third thirds (both $p < .001$), indicating a progressive reorientation rather than an instant response to the explanation's opening.

For dual nature theory, this asymmetry is notable. The two perspectives do not appear to be symmetrically positioned in explanatory dialogue. Architecture content, even when dominant, does not pull EEs who have been exposed to Relevance back toward a purely structural orientation. Relevance, once introduced, reshapes the co-constructive dynamic in a way that Architecture alone does not. This is consistent with the cognitive preference for functional reasoning discussed in the Related Work \citep[cf.][]{kelemen_function_1999,lombrozo_mechanistic_2019}. From a practical standpoint, the finding implies that Relevance-oriented engagement may require active scaffolding from the EX, as it does not emerge spontaneously from Architecture-focused explanation. Achieving integrated understanding during the co-constructive process may therefore depend on the EX actively introducing Relevance perspectives that would not surface on their own. A more detailed analysis of EE temporal dynamics, including speaker-level interaction patterns across conditions, is planned for future work.

\subsection{Practical Implications}

\textbf{For Education.} The equivalence of the A and R conditions suggests that technology educators need not worry about choosing the ``right'' content focus for promoting comprehension, as both perspectives achieve comparable results. However, if the goal is to give students more than comprehension---the ability to use, evaluate, and adapt technical tools (i.e., agency)---then explicitly connecting Architecture and Relevance should be preferred over teaching one at the expense of the other. This aligns with a broader disciplinary observation. \citet{tedre_three_2013} identify three traditions that have shaped computing and its teaching: a theoretical tradition concerned with formal structures, an engineering tradition concerned with producing working implementations, and a scientific tradition concerned with empirical investigation. The theoretical tradition privileges propositional knowledge (know-that), while the engineering tradition uniquely integrates propositional and procedural knowledge (know-that and know-how). Our findings map onto this distinction: focused explanations produce equivalent comprehension regardless of content emphasis, but only integrated explanations produce enabledness. In \citeauthor{tedre_three_2013}'s terms, single-perspective explanations may produce theoretical-tradition-style understanding, whereas integration fosters the engineering tradition's epistemic stance of connecting what an artifact is for with how it is built. If computing education curricula remain anchored primarily in the theoretical tradition, as \citeauthor{tedre_three_2013} suggest is common, the kind of integration that our data show to be necessary for enabledness may be systematically undertaught. Curricula should incorporate explicit bridging exercises that require students to trace connections between implementation and purpose, potentially through activities like ``explanation generation'' where students must explain artifacts to others. The goal should not merely be understanding \textit{that} an artifact works or \textit{how} it works, but developing the capacity to use this knowledge productively through their integration.

\medskip
\noindent
\textbf{For XAI Design.} \citet{miller_explanation_2019} argues that XAI research has been too focused on the researcher's perspective, neglecting how people actually process and use explanations. Our findings give empirical shape to this critique. Most XAI evaluation focuses on whether users comprehend the system's reasoning, but our data show that comprehension is largely insensitive to how Architecture and Relevance are balanced---what integration specifically enhances is enabledness, the capacity to act with and upon the system. This suggests that XAI evaluation should shift from measuring explanation comprehension toward assessing whether explanations enable users to make informed decisions, a point \citet{sokol_one_2020} raise when arguing that explanation quality cannot be assessed independently of what the user needs to accomplish. The exploratory ratchet dynamic observed in our data suggests a tentative design principle: introducing Relevance content early, even briefly, may reshape user engagement with subsequent Architecture content without requiring a perfectly balanced presentation throughout. Rather than engineering a single optimal explanation, systems could front-load a Relevance frame (e.g., what the decision means for the user and what actions are available) before presenting architectural detail (e.g., feature weights or decision paths). Finally, the persistent Architecture-dominant baseline across all conditions constitutes a warning for systems designed by technical practitioners. Engineer-authored explanations will likely reproduce the mechanistic stance default observed in our data unless Relevance integration is built in as a design constraint rather than added as an afterthought.

\subsection{Limitations}

Several limitations should be noted when interpreting these findings. Regarding the understanding measure, the questionnaire was developed through a three-phase process involving interview-based item generation and two rounds of exploratory factor analysis on independent samples ($N = 157$, $N = 134$), documented in an OSF repository.\footnote{\url{https://osf.io/w39dc/overview?view_only=517962d50d314c529b6d0aafe20cf650}} One Relevance item (R13) was excluded post-hoc on substantive grounds (both response options were strategically defensible). The instrument showed modest internal consistency in its final application ($\alpha = .49$ for enabledness), expected given the content heterogeneity and ceiling effects introduced by effective face-to-face explanations. Since measurement error attenuates effects, the significant enabledness result is best read as a conservative estimate. Across both EFA phases, Architecture and Relevance items consistently loaded on shared factors rather than separating along the theoretical distinction, supporting the decision to aggregate them into a single comprehension score---though this aggregation may mask the condition-specific effects that H1 and H2 predicted. Future work with greater item difficulty, feasible with a more complex explanandum, could improve subscale separability.

The total sample ($N$ = 104) provides moderate statistical power for detecting medium-sized effects. Two participants who failed the Instructional Manipulation Check \citep{oppenheimer_instructional_2009} were excluded from the understanding analyses. When included, Cronbach's $\alpha$ inflates from .60 to .80 and distorts normality in the AR group ($W$ dropping from .87 to .69), both consistent with patterned responding. When retained, the enabledness effect is no longer significant ($F(1, 105) = 2.59$, $p = .111$). We report both analyses for transparency.

The controlled experimental setting with confederate EXs maximized internal validity but necessarily reduced ecological validity compared to, for example, natural teaching contexts. Naturalness ratings, which were uniformly high and did not differ across conditions, provide some reassurance that the confederate design did not introduce perceptible artificiality, though we acknowledge that ratings cannot capture all aspects of ecological validity.

Explanation duration varied significantly across conditions, with Architecture-focused explanations being substantially shorter than the other two conditions. While this reflects a genuine property of the conditions (less content to cover when one perspective is minimized) it means that duration is confounded with condition. That said, the AR and R condition produced comparably long explanations  ($p = .190$), without the enabledness benefit in the AR condition, which argues against a pure time-on-task interpretation.

The physical absence of Quarto! during the explanation was a deliberate design choice intended to encourage explicit verbalization and to support future gesture analysis. However, it also means that the explanations unfolded without the support that a physical artifact provides. In naturalistic settings, explainers frequently point to, manipulate, or demonstrate with the artifact itself, which may alter both the structure of the explanation and the understanding it produces. The present findings therefore characterize explanation dynamics in a ``black box'' scenario and may not generalize directly to contexts where the artifact is available for joint inspection. This potentially influences the amount of Architecture related utterances in the explanation, as the material of the game needs description. With a present game we expect less Architecture utterances and therefore the balance might shift throughout conditions. Future work should assess how and if that would influence the results of the temporal patterns. 

Finally, the explanandum itself constrains generalizability. This study was designed as foundational research: the goal was to test theoretically grounded hypotheses about the effect of dual-nature emphasis on understanding under conditions that minimize confounding variables, using an explanandum accessible enough to recruit a broad participant pool without requiring domain-specific prior knowledge. Quarto! serves this purpose well, exhibiting clear dual-nature characteristics that make it suited for isolating core explanatory mechanisms. However, it is a board game of limited complexity. Its Architecture can be grasped quickly, and its Relevance, though strategically rich, does not approach the contested, multi-stakeholder character of Relevance in complex digital artifacts such as AI systems or algorithmic decision-making tools. Whether the integration advantage on enabledness holds---or becomes even more pronounced---for artifacts where the Architecture is partially opaque and the Relevance is socially contested remains an open and important question.

Future research should examine whether integrated explanation strategies transfer to complex digital artifacts. Replication with larger samples and improved measurement instruments, particularly for the enabledness construct and with separable comprehension subscales, would strengthen confidence in the present findings. Investigating how learners can be trained to generate bridging connections autonomously, rather than relying on explainer-provided bridges, represents a promising direction for educational practice.

\section{Conclusion}

This study demonstrates that the way EXs address the dual nature of technical artifacts affects understanding outcomes, though the pattern of effects differed from our original predictions. We had hypothesized that each focused condition (A-focus, R-focus) would produce a characteristic understanding profile (H1, H2) and that integrated explanations would produce the highest levels of both comprehension and enabledness (H3). Instead, the data revealed that the content focus of an explanation, whether predominantly on Architecture or predominantly on Relevance, had no detectable effect on any understanding outcome. What mattered was whether the explanation integrated both perspectives: the AR condition specifically enhanced enabledness (the ability to apply understanding in practical contexts) while comprehension was comparable to the two other focused conditions. However, it is crucial to note that the null effect on H1 and H2 may be an artifact of an explanandum that was undercomplicated or an artifact of the inadequacy of the understanding measurement to capture A and R comprehension in isolation.

But this pattern of partial hypothesis support is, on reflection, very informative. The null effect of content focus (disconfirming H1 and H2), given it really can be confirmed with more complex artifacts in future research, suggests that learners are more resilient to variation in explanation content than we assumed. The integrated focus' advantage on enabledness (partially confirming H3) suggests that what integration provides is not just knowledge but the capacity to apply Architecture and Relevance comprehension in action. This is consistent with claims around the dual nature framework's stating that genuine understanding of technical artifacts requires bridging Architecture and Relevance \citep{schulte_framework_2018,winkelnkemper_ariadne_2024,deridder_reconstructing_2007}, while specifying that this bridging is specifically necessary for enabledness rather than for comprehension.

For the broader goals of the research project, these results provide empirical grounding for the claim that promoting agency, the ability to understand, evaluate, and adapt technical artifacts to one's needs, requires more than conveying knowledge about either Architecture or Relevance in isolation. It requires helping learners build bridges between these perspectives.

\section*{Declarations}

\textbf{Funding}
\\
\noindent
This research was funded by REMOVED FOR REVIEW %the Deutsche Forschungsgemeinschaft (DFG, German Research Foundation): TRR 318/1 2021 -- 438445824.

\noindent
\textbf{Ethics Statement}
\noindent
\\
This study was approved by REMOVED FOR REVIEW %the Paderborn University 
Ethics Board. 
All subjects participated voluntarily and provided written informed consent before the studies. 

\noindent
\textbf{Acknowledgements}
\\\noindent
We thank our colleagues for feedback on the first drafts, as well as all research assistants for their support during coding, analyses, and data acquisition.

\textbf{Disclosure}
\noindent
\\
AI-assisted tools were used for checking linguistic correctness (LanguageTool, Claude), and during data analysis in Python (Copilot in VSCode). All recommendations were systematically double-checked to meet a high quality standard. 

\noindent
\textbf{Conflict of Interest}
\\\noindent
On behalf of all authors, the corresponding author states that there is no conflict of interest.

%\begin{appendices}

%\section{Section title of first appendix}\label{secA1}

%An appendix contains supplementary information that is not an essential part of the text itself but which may be helpful in providing a more comprehensive understanding of the research problem or it is information that is too cumbersome to be included in the body of the paper.

%%=============================================%%
%% For submissions to Nature Portfolio Journals %%
%% please use the heading ``Extended Data''.   %%
%%=============================================%%

%%=============================================================%%
%% Sample for another appendix section			       %%
%%=============================================================%%

%% \section{Example of another appendix section}\label{secA2}%
%% Appendices may be used for helpful, supporting or essential material that would otherwise 
%% clutter, break up or be distracting to the text. Appendices can consist of sections, figures, 
%% tables and equations etc.

%\end{appendices}

%%===========================================================================================%%
%% If you are submitting to one of the Nature Portfolio journals, using the eJP submission   %%
%% system, please include the references within the manuscript file itself. You may do this  %%
%% by copying the reference list from your .bbl file, paste it into the main manuscript .tex %%
%% file, and delete the associated \verb+\bibliography+ commands.                            %%
%%===========================================================================================%%

\bibliography{latex/references}% common bib file

@article{terfloth_navigating_inpress,
	title = {Navigating the {Dual} {Nature}: {Do} {Explainers} {Adapt} to {Explainee} {Interests} {When} {Explaining} {Technical} {Artifacts}},
	language = {en},
	journal = {International Journal of Technology and Design Education},
	publisher = {Springer},
	author = {Terfloth, Lutz and Buhl, Heike M and Lohmer, Vivien and Kern, Friederike and Schaffer, Michael and Schulte, Carsten},
}

@book{streeck_gesturecraft_2009,
	address = {Amsterdam},
	series = {Gesture {Studies}},
	title = {Gesturecraft: {The} manu-facture of meaning},
	volume = {2},
	isbn = {978-90-272-2842-0 978-90-272-2846-8 978-90-272-8982-7},
	shorttitle = {Gesturecraft},
	url = {http://www.jbe-platform.com/content/books/9789027289827},
	doi = {10.1075/gs.2},
	language = {en},
	urldate = {2026-03-25},
	publisher = {John Benjamins Publishing Company},
	author = {Streeck, Jürgen},
	month = apr,
	year = {2009},
}

@book{fisher_exploring_2023,
	title = {Exploring the {Semantic} {Dialogue} {Patterns} of {Explanations} - a {Case} {Study} of {Game} {Explanations}},
	author = {Fisher, Josephine and Robrecht, Amelie and Kopp, Stefan and Rohlfing, Katharina},
	month = aug,
	year = {2023},
}

@incollection{tedre_three_2013,
	title = {Three computing traditions in school computing education},
	booktitle = {Improving computer science education},
	publisher = {Routledge New York, NY and London},
	author = {Tedre, Matti and Apiola, Mikko},
	editor = {Kadijevich, D. M. and Angeli, C. and Schulte, C.},
	year = {2013},
	note = {00013},
	pages = {100--116},
}

@incollection{lombrozo_mechanistic_2019,
	title = {Mechanistic versus {Functional} {Understanding}},
	isbn = {978-0-19-086098-1},
	doi = {10.1093/oso/9780190860974.003.0011},
	language = {en},
	booktitle = {Varieties of {Understanding}: {New} {Perspectives} from {Philosophy}, {Psychology}, and {Theology}},
	publisher = {Oxford University Press},
	author = {Lombrozo, Tania and Wilkenfeld, Daniel},
	editor = {Grimm, Stephen R.},
	month = aug,
	year = {2019},
}

@inproceedings{schulte_framework_2018,
	address = {Koli, Finland},
	title = {A {Framework} for {Computing} {Education}: {Hybrid} {Interaction} {System}: {The} need for a bigger picture in computing education.},
	volume = {18},
	doi = {10.1145/3279720.3279733},
	booktitle = {18th {Koli} {Calling} {International} {Conference} on {Computing} {Education} {Research} ({Koli} {Calling} ’18)},
	publisher = {ACM},
	author = {Schulte, Carsten and Budde, Lea},
	month = nov,
	year = {2018},
	pages = {10},
}

@incollection{pennington_comprehension_1987,
	title = {Comprehension strategies in programming},
	isbn = {0-89391-461-4},
	url = {http://portal.acm.org/citation.cfm?id=54968.54975},
	doi = {10.5555/54968.54975},
	booktitle = {Empirical studies of programmers: second workshop},
	publisher = {Ablex Publishing Corp.},
	author = {Pennington, Nancy},
	editor = {Olson, Gary M. and Sheppard, Sylvia and Soloway, Elliot},
	year = {1987},
	pages = {100--113},
}

@book{deridder_reconstructing_2007,
	title = {Reconstructing design, explaining artifacts: philosophical reflections on the design and explanation of technical artifacts},
	isbn = {978-90-90-21903-5},
	url = {https://repository.tudelft.nl/record/uuid:d67f6fe7-59c5-4357-903e-e3c3891e2721},
	author = {de Ridder, Jeroen},
	year = {2007},
}

@article{spohrer_goal_1985,
	title = {A {Goal}/{Plan} {Analysis} of {Buggy} {Pascal} {Programs}},
	volume = {1},
	issn = {0737-0024, 1532-7051},
	url = {http://www.tandfonline.com/doi/abs/10.1207/s15327051hci0102_4},
	doi = {10.1207/s15327051hci0102_4},
	language = {en},
	number = {2},
	urldate = {2020-12-06},
	journal = {Human–Computer Interaction},
	publisher = {Taylor \& Francis},
	author = {Spohrer, James C. and Soloway, Elliot and Pope, Edgar},
	month = jun,
	year = {1985},
	pages = {163--207},
}

@article{kroes_engineering_2010,
	title = {Engineering and the dual nature of technical artefacts},
	volume = {34},
	doi = {10.1093/cje/bep019},
	number = {1},
	journal = {Cambridge journal of economics},
	publisher = {Oxford University Press},
	author = {Kroes, Peter},
	year = {2010},
	pages = {51--62},
}

@inproceedings{register_learning_2020,
	address = {Virtual Event New Zealand},
	title = {Learning {Machine} {Learning} with {Personal} {Data} {Helps} {Stakeholders} {Ground} {Advocacy} {Arguments} in {Model} {Mechanics}},
	isbn = {978-1-4503-7092-9},
	url = {https://dl.acm.org/doi/10.1145/3372782.3406252},
	doi = {10.1145/3372782.3406252},
	language = {en},
	urldate = {2020-12-07},
	booktitle = {Proceedings of the 2020 {ACM} {Conference} on {International} {Computing} {Education} {Research}},
	publisher = {ACM},
	author = {Register, Yim and Ko, Amy J.},
	month = aug,
	year = {2020},
	pages = {67--78},
}

@article{keil_explanation_2006,
	title = {Explanation and understanding},
	volume = {57},
	url = {https://doi.org/10.1146/annurev.psych.57.102904.190100},
	doi = {10.1146/annurev.psych.57.102904.190100},
	number = {1},
	journal = {Annual Review of Psychology},
	author = {Keil, Frank C.},
	year = {2006},
	pages = {227--254},
}

@inproceedings{schulte_block_2008,
	address = {Sydney, Australia},
	series = {{ICER} '08},
	title = {Block {Model} - an {Educational} {Model} of {Program} {Comprehension} as a {Tool} for a {Scholarly} {Approach} to {Teaching}},
	isbn = {978-1-60558-216-0},
	url = {http://doi.acm.org/10.1145/1404520.1404535},
	doi = {10.1145/1404520.1404535},
	booktitle = {Proceeding of the {Fourth} international {Workshop} on {Computing} {Education} {Research}},
	publisher = {ACM},
	author = {Schulte, Carsten},
	year = {2008},
	pages = {149--160},
}

@article{nuckles_information_2006,
	title = {Information {About} a {Layperson}'s {Knowledge} {Supports} {Experts} in {Giving} {Effective} and {Efficient} {Online} {Advice} to {Laypersons}.},
	volume = {11},
	doi = {10.1037/1076-898X.11.4.219},
	journal = {Journal of experimental psychology. Applied},
	author = {Nückles, Matthias and Wittwer, Jörg and Renkl, Alexander},
	month = jan,
	year = {2006},
	pages = {219--36},
}

@article{wittwer_can_2010,
	title = {Can {Tutors} {Be} {Supported} in {Giving} {Effective} {Explanations}?},
	volume = {102},
	doi = {10.1037/a0016727},
	journal = {Journal of Educational Psychology},
	author = {Wittwer, Jörg and Nückles, Matthias and Landmann, Nina and Renkl, Alexander},
	month = feb,
	year = {2010},
	pages = {74--89},
}

@article{rohlfing_explanation_2021,
	title = {Explanation as a {Social} {Practice}: {Toward} a {Conceptual} {Framework} for the {Social} {Design} of {AI} {Systems}},
	volume = {13},
	issn = {2379-8920, 2379-8939},
	shorttitle = {Explanation as a {Social} {Practice}},
	url = {https://ieeexplore.ieee.org/document/9292993/},
	doi = {10.1109/TCDS.2020.3044366},
	language = {en},
	number = {3},
	urldate = {2021-10-25},
	journal = {IEEE Transactions on Cognitive and Developmental Systems},
	author = {Rohlfing, Katharina J. and Cimiano, Philipp and Scharlau, Ingrid and Matzner, Tobias and Buhl, Heike M. and Buschmeier, Hendrik and Esposito, Elena and Grimminger, Angela and Hammer, Barbara and Hab-Umbach, Reinhold and Horwath, Ilona and Hullermeier, Eyke and Kern, Friederike and Kopp, Stefan and Thommes, Kirsten and Ngonga Ngomo, Axel-Cyrille and Schulte, Carsten and Wachsmuth, Henning and Wagner, Petra and Wrede, Britta},
	month = sep,
	year = {2021},
	pages = {717--728},
}

@article{graesser_collaborative_1995,
	title = {Collaborative dialogue patterns in naturalistic one-to-one tutoring},
	volume = {9},
	copyright = {Copyright © 1995 John Wiley \& Sons, Ltd},
	issn = {1099-0720},
	url = {https://onlinelibrary.wiley.com/doi/abs/10.1002/acp.2350090604},
	doi = {10.1002/acp.2350090604},
	language = {en},
	number = {6},
	urldate = {2025-07-02},
	journal = {Applied Cognitive Psychology},
	author = {Graesser, Arthur C. and Person, Natalie K. and Magliano, Joseph P.},
	year = {1995},
	pages = {495--522},
}

@incollection{dennett_intentional_2009,
	title = {Intentional {Systems} {Theory}},
	isbn = {978-0-19-926261-8},
	url = {https://doi.org/10.1093/oxfordhb/9780199262618.003.0020},
	doi = {10.1093/oxfordhb/9780199262618.003.0020},
	booktitle = {The {Oxford} {Handbook} of {Philosophy} of {Mind}},
	publisher = {Oxford University Press},
	author = {Dennett, Daniel},
	editor = {Beckermann, Ansgar and McLaughlin, Brian P. and Walter, Sven},
	month = jan,
	year = {2009},
	pages = {339--350},
}

@article{chi_icap_2014,
	title = {The {ICAP} {Framework}: {Linking} {Cognitive} {Engagement} to {Active} {Learning} {Outcomes}},
	volume = {49},
	issn = {0046-1520},
	shorttitle = {The {ICAP} {Framework}},
	url = {https://doi.org/10.1080/00461520.2014.965823},
	doi = {10.1080/00461520.2014.965823},
	number = {4},
	urldate = {2022-08-01},
	journal = {Educational Psychologist},
	publisher = {Routledge},
	author = {Chi, Michelene T. H. and Wylie, Ruth},
	month = oct,
	year = {2014},
	pages = {219--243},
}

@inproceedings{schulte_introduction_2010,
	address = {New York, NY, USA},
	series = {{ITiCSE}-{WGR} '10},
	title = {An introduction to program comprehension for computer science educators},
	isbn = {978-1-4503-0677-5},
	url = {https://doi.org/10.1145/1971681.1971687},
	doi = {10.1145/1971681.1971687},
	urldate = {2022-11-10},
	booktitle = {Proceedings of the 2010 {ITiCSE} working group reports},
	publisher = {Association for Computing Machinery},
	author = {Schulte, Carsten and Clear, Tony and Taherkhani, Ahmad and Busjahn, Teresa and Paterson, James H.},
	month = jun,
	year = {2010},
	pages = {65--86},
}

@article{miyake_constructive_1986,
	title = {Constructive {Interaction} and the {Iterative} {Process} of {Understanding}},
	volume = {10},
	issn = {0364-0213, 1551-6709},
	url = {https://onlinelibrary.wiley.com/doi/10.1207/s15516709cog1002_2},
	doi = {10.1207/s15516709cog1002_2},
	language = {en},
	number = {2},
	urldate = {2024-03-07},
	journal = {Cognitive Science},
	author = {Miyake, Naomi},
	month = apr,
	year = {1986},
	pages = {151--177},
}

@article{foxtree_listening_1999,
	address = {US},
	title = {Listening in on monologues and dialogues},
	volume = {27},
	issn = {1532-6950},
	doi = {10.1080/01638539909545049},
	number = {1},
	journal = {Discourse Processes},
	publisher = {Lawrence Erlbaum},
	author = {Fox Tree, Jean E.},
	year = {1999},
	pages = {35--53},
}

@article{buschmeier_forms_2025,
	title = {Forms of understanding for {XAI}-{Explanations}},
	volume = {94},
	issn = {1389-0417},
	url = {https://www.sciencedirect.com/science/article/pii/S1389041725000993},
	doi = {10.1016/j.cogsys.2025.101419},
	urldate = {2026-02-19},
	journal = {Cognitive Systems Research},
	author = {Buschmeier, Hendrik and Buhl, Heike M. and Kern, Friederike and Grimminger, Angela and Beierling, Helen and Fisher, Josephine and Groß, André and Horwath, Ilona and Klowait, Nils and Lazarov, Stefan and Lenke, Michael and Lohmer, Vivien and Rohlfing, Katharina and Scharlau, Ingrid and Singh, Amit and Terfloth, Lutz and Vollmer, Anna-Lisa and Wang, Yu and Wilmes, Annedore and Wrede, Britta},
	month = dec,
	year = {2025},
	pages = {101419},
}

@article{mccarthy_right_2023,
	title = {A right way to explain? {Function}, mechanism, and the order of explanations},
	volume = {238},
	issn = {00100277},
	shorttitle = {A right way to explain?},
	url = {https://linkinghub.elsevier.com/retrieve/pii/S0010027723001282},
	doi = {10.1016/j.cognition.2023.105494},
	language = {en},
	urldate = {2026-02-18},
	journal = {Cognition},
	author = {McCarthy, Amanda M. and Keil, Frank C.},
	month = sep,
	year = {2023},
	pages = {105494},
}

@article{oppenheimer_instructional_2009,
	title = {Instructional manipulation checks: {Detecting} satisficing to increase statistical power},
	volume = {45},
	issn = {0022-1031},
	shorttitle = {Instructional manipulation checks},
	url = {https://www.sciencedirect.com/science/article/pii/S0022103109000766},
	doi = {10.1016/j.jesp.2009.03.009},
	number = {4},
	urldate = {2026-02-18},
	journal = {Journal of Experimental Social Psychology},
	author = {Oppenheimer, Daniel M. and Meyvis, Tom and Davidenko, Nicolas},
	month = jul,
	year = {2009},
	pages = {867--872},
}

@inproceedings{wittenburg_elan_2006,
	address = {Genoa, Italy},
	title = {{ELAN} : a professional framework for multimodality research},
	shorttitle = {{ELAN}},
	url = {https://hdl.handle.net/11858/00-001M-0000-0013-1E7E-4},
	urldate = {2025-08-05},
	author = {Wittenburg, Peter and Brugman, Hennie and Russel, Albert and Klassmann, Alex and Sloetjes, Han},
	year = {2006},
	pages = {1556--1559},
}

@article{terfloth_transcription_2025,
	title = {Transcription in {Computing} {Education} {Research}: {A} {Review} and {Recommendations}},
	issn = {1648-5831, 2335-8971},
	shorttitle = {Transcription in {Computing} {Education} {Research}},
	url = {https://www.infedu.vu.lt/journal/INFEDU/article/805},
	doi = {10.15388/infedu.2025.09},
	language = {en},
	urldate = {2025-05-15},
	journal = {Informatics in Education},
	publisher = {Vilnius University Institute of Data Science and Digital Technologies},
	author = {Terfloth, Lutz and Lohmer, Vivien and Kern, Friederike and Schulte, Carsten},
	month = apr,
	year = {2025},
}

@incollection{terfloth_adding_2023,
	address = {Cham},
	title = {Adding {Why} to {What}? {Analyses} of an {Everyday} {Explanation}},
	volume = {1903},
	isbn = {978-3-031-44069-4 978-3-031-44070-0},
	shorttitle = {Adding {Why} to {What}?},
	url = {https://link.springer.com/10.1007/978-3-031-44070-0_13},
	doi = {10.1007/978-3-031-44070-0_13},
	language = {en},
	urldate = {2024-02-27},
	booktitle = {Explainable {Artificial} {Intelligence}},
	publisher = {Springer Nature},
	author = {Terfloth, Lutz and Schaffer, Michael and Buhl, Heike M. and Schulte, Carsten},
	editor = {Longo, Luca},
	year = {2023},
	pages = {256--279},
}

@article{frederik_teaching_2011,
	title = {Teaching and learning the nature of technical artifacts},
	volume = {21},
	copyright = {http://creativecommons.org/licenses/by-nc/2.0},
	issn = {0957-7572, 1573-1804},
	url = {http://link.springer.com/10.1007/s10798-010-9119-3},
	doi = {10.1007/s10798-010-9119-3},
	language = {en},
	number = {3},
	urldate = {2025-07-28},
	journal = {International Journal of Technology and Design Education},
	publisher = {Springer Science and Business Media LLC},
	author = {Frederik, Ineke and Sonneveld, Wim and De Vries, Marc J.},
	month = aug,
	year = {2011},
	pages = {277--290},
}

@book{mcneill_hand_1992,
	address = {Chicago},
	title = {Hand and mind: what gestures reveal about thought},
	isbn = {978-0-226-56132-5},
	shorttitle = {Hand and mind},
	publisher = {University of Chicago Press},
	author = {McNeill, David},
	year = {1992},
}

@incollection{clark_grounding_1991,
	address = {Washington, DC, US},
	title = {Grounding in communication},
	isbn = {978-1-55798-121-9},
	doi = {10.1037/10096-006},
	booktitle = {Perspectives on socially shared cognition},
	publisher = {American Psychological Association},
	author = {Clark, Herbert H. and Brennan, Susan E.},
	year = {1991},
	pages = {127--149},
}

@article{winkelnkemper_ariadne_2024,
	title = {{ARIadne} – {An} {Explanation} {Model} for {Digital} {Artefacts}},
	volume = {23},
	issn = {1648-5831, 2335-8971},
	url = {https://infedu.vu.lt/journal/INFEDU/article/772},
	doi = {10.15388/infedu.2024.09},
	language = {en},
	number = {2},
	urldate = {2024-07-11},
	journal = {Informatics in Education},
	publisher = {Vilnius University Institute of Data Science and Digital Technologies},
	author = {Winkelnkemper, Felix and Höper, Lukas and Schulte, Carsten},
	month = jun,
	year = {2024},
	pages = {479--505},
}

@incollection{kuckartz_transcribing_2019,
	address = {Cham},
	title = {Transcribing {Audio} and {Video} {Recordings}},
	isbn = {978-3-030-15671-8},
	url = {https://doi.org/10.1007/978-3-030-15671-8_4},
	doi = {10.1007/978-3-030-15671-8_4},
	language = {en},
	urldate = {2024-02-12},
	booktitle = {Analyzing {Qualitative} {Data} with {MAXQDA}: {Text}, {Audio}, and {Video}},
	publisher = {Springer International Publishing},
	author = {Kuckartz, Udo and Rädiker, Stefan},
	editor = {Kuckartz, Udo and Rädiker, Stefan},
	year = {2019},
	pages = {41--49},
}

@article{miller_explanation_2019,
	title = {Explanation in artificial intelligence: {Insights} from the social sciences},
	volume = {267},
	issn = {0004-3702},
	shorttitle = {Explanation in artificial intelligence},
	url = {https://www.sciencedirect.com/science/article/pii/S0004370218305988},
	doi = {10.1016/j.artint.2018.07.007},
	language = {en},
	urldate = {2022-10-31},
	journal = {Artificial Intelligence},
	author = {Miller, Tim},
	month = feb,
	year = {2019},
	pages = {1--38},
}

@article{selting_gesprachsanalytisches_2009,
	title = {Gesprächsanalytisches {Transkriptionssystem} ({GAT} 2)},
	volume = {10},
	journal = {Gesprächsforsch},
	author = {Selting, Margret and Auer, Peter and Barth-Weingarten, D. and Bergmann, Jörg and Bergmann, Pia and Birkner, Karin and Couper-Kuhlen, Elizabeth and Deppermann, Arnulf and Gilles, Peter and Günthner, Susanne and Hartung, Martin and Kern, Friederike and Mertzlufft, C. and Meyer, Christian and Morek, Miriam and Oberzaucher, F. and Peters, Jörg and Quasthoff, U. and Schütte, Wilfried and Uhmann, Susanne},
	month = jan,
	year = {2009},
	pages = {152--183},
}

@article{soloway_learning_1986,
	title = {Learning to program = learning to construct mechanisms and explanations},
	volume = {29},
	issn = {0001-0782},
	url = {https://doi.org/10.1145/6592.6594},
	doi = {10.1145/6592.6594},
	number = {9},
	urldate = {2022-06-02},
	journal = {Communications of the ACM},
	author = {Soloway, E.},
	month = sep,
	year = {1986},
	pages = {850--858},
}

@article{hoper_data_2023,
	title = {The data awareness framework as part of data literacies in {K}-12 education},
	volume = {125},
	issn = {2398-5348},
	url = {https://doi.org/10.1108/ILS-06-2023-0075},
	doi = {10.1108/ILS-06-2023-0075},
	number = {7-8},
	urldate = {2025-09-12},
	journal = {Information and Learning Sciences},
	author = {Höper, Lukas and Schulte, Carsten},
	month = dec,
	year = {2023},
	pages = {491--512},
}

@article{sokol_one_2020,
	title = {One {Explanation} {Does} {Not} {Fit} {All}},
	volume = {34},
	issn = {1610-1987},
	url = {https://doi.org/10.1007/s13218-020-00637-y},
	doi = {10.1007/s13218-020-00637-y},
	language = {en},
	number = {2},
	urldate = {2025-09-12},
	journal = {KI - Künstliche Intelligenz},
	author = {Sokol, Kacper and Flach, Peter},
	month = jun,
	year = {2020},
	pages = {235--250},
}

@incollection{kintsch_role_1991,
	series = {Text and {Text} {Processing}},
	title = {The {Role} of {Knowledge} in {Discourse} {Comprehension}: {A} {Construction}-{Integration} {Model}},
	volume = {79},
	shorttitle = {The {Role} of {Knowledge} in {Discourse} {Comprehension}},
	url = {https://www.sciencedirect.com/science/article/pii/S0166411508615514},
	doi = {10.1016/S0166-4115(08)61551-4},
	urldate = {2025-07-15},
	booktitle = {Advances in {Psychology}},
	publisher = {North-Holland},
	author = {Kintsch, Walter},
	editor = {Stelmach, G. E. and Vroon, P. A.},
	month = jan,
	year = {1991},
	pages = {107--153},
}

@article{kelemen_function_1999,
	title = {Function, goals and intention: children’s teleological reasoning about objects},
	volume = {3},
	issn = {13646613},
	shorttitle = {Function, goals and intention},
	url = {https://linkinghub.elsevier.com/retrieve/pii/S1364661399014023},
	doi = {10.1016/S1364-6613(99)01402-3},
	language = {en},
	number = {12},
	urldate = {2022-09-21},
	journal = {Trends in Cognitive Sciences},
	author = {Kelemen, Deborah},
	month = dec,
	year = {1999},
	pages = {461--468},
}

@inproceedings{kallia_search_2023,
	address = {Chicago IL USA},
	title = {The {Search} for {Meaning}: {Inferential} {Strategic} {Reading} {Comprehension} in {Programming}},
	isbn = {978-1-4503-9976-0},
	shorttitle = {The {Search} for {Meaning}},
	url = {https://dl.acm.org/doi/10.1145/3568813.3600135},
	doi = {10.1145/3568813.3600135},
	language = {en},
	urldate = {2023-10-31},
	booktitle = {Proceedings of the 2023 {ACM} {Conference} on {International} {Computing} {Education} {Research} {V}.1},
	publisher = {ACM},
	author = {Kallia, Maria},
	month = aug,
	year = {2023},
	pages = {1--14},
}
%% if required, the content of .bbl file can be included here once bbl is generated
%%\input sn-article.bbl

\end{document}